\documentclass{sig-alt-full}

\usepackage[utf8]{inputenc}
\usepackage{amssymb,amsmath}
\usepackage{theorem}
\usepackage{graphics}
\usepackage{amsfonts}
\usepackage{mathrsfs}
\usepackage{amscd}
\usepackage{color}
\usepackage{url}
\usepackage[plainpages=false,pdfpagelabels,colorlinks=true,citecolor=blue,hypertexnames=false]{hyperref}
\usepackage{alltt}
\usepackage{bbm}
\usepackage{pst-grad,auto-pst-pdf}

\def\gathen#1{{#1}}
\def\hoeven#1{{#1}}

\def\myproof{\noindent{\sc Proof.}~}
\def\foorp{\hfill$\square$}

\def\MM {\ensuremath{\mathsf{MM}}}
\def\C {\ensuremath{\mathsf{C}}}
\def\F {\ensuremath{\mathbb{F}}}
\def\K {\ensuremath{\mathbb{K}}}
\newcommand{\x}{x}
\newcommand{\Dx}{\partial}
\newcommand{\cN}{ {\cal N}}
\newcommand{\pa} { \partial}
\newcommand{\ie}{{\it i.e.}}
\newcommand{\rem }{ {\rm rem}}
\newcommand{\rank}   { {\rm rank} }
\newcommand{\vu} { {\bf u}}
\newcommand{\vv}{ {\bf v}}

\newcommand{\ord} { {\rm ord}}
\newcommand{\lclm} { {\rm LCLM}}
\newcommand{\bigOsoft}{\widetilde{{O}}}

\newtheorem{theorem}{Theorem}

\newtheorem{cor}[theorem]{Corollary}
\newtheorem{lemma}[theorem]{Lemma}

\def\qed{\hfil {\vrule height5pt width2pt depth2pt}}

\begin{document}

\title{Fast Computation of Common Left Multiples \\
of Linear Ordinary Differential Operators%
\titlenote{\small %
We warmly thank the referees for their very helpful comments.
---
This work was supported in part by the MSR--INRIA Joint Centre,
and by two NSFC grants (91118001 and 60821002/F02). 
\vspace{-36pt}}}
\newfont{\authfntsmall}{phvr at 11pt}
\newfont{\eaddfntsmall}{phvr at 9pt}
\def\more-auths{%
\end{tabular}
\begin{tabular}{c}}
\numberofauthors{2}
\author{
\alignauthor {\authfntsmall Alin Bostan, Fr\'ed\'eric Chyzak, Bruno Salvy}\\
\affaddr{Algorithms Project, INRIA (France)}\\
\email{\eaddfntsmall{\{alin.bostan,frederic.chyzak,bruno.salvy\}@inria.fr}}
\alignauthor {\authfntsmall Ziming Li}\\
\affaddr{KLMM and AMSS (China)}\\
\email{\eaddfntsmall{zmli@mmrc.iss.ac.cn}}}

\maketitle
\begin{abstract}
We study tight bounds and fast algorithms
for LCLMs of 
\emph{several\/} 
linear differential operators with polynomial coefficients. We analyse the
arithmetic complexity of existing algorithms for LCLMs, as well as the size of
their outputs. We propose a new algorithm that recasts the LCLM computation in
a linear algebra problem on a polynomial matrix. This algorithm yields sharp
bounds on the coefficient degrees of the LCLM, improving by one order of
magnitude the best bounds obtained using previous algorithms. The complexity
of the new algorithm is almost optimal, in the sense that it nearly matches
the arithmetic size of the output. \end{abstract}

\vspace{1mm}
\noindent
{\bf Categories and Subject Descriptors:} \\
\noindent I.1.2 [{\bf Computing Methodologies}]: Symbolic and
Algebraic Manipulations --- \emph{Algebraic Algorithms}

\vspace{1mm}
\noindent {\bf General Terms:} Algorithms, Theory.

\vspace{1mm}


\section{Introduction} 
The complexity of operations in the polynomial ring $\K[\x]$ over a
field~$\K$ has been intensively studied in the computer algebra literature.
It is well established that polynomial multiplication is a \emph{commutative
complexity yardstick}, in the sense that the complexity of operations
in~$\K[\x]$ can be expressed in terms of the cost of multiplication, and for
most of them, in a quasi-linear way.

Linear differential operators in the derivation $\Dx =
\frac{\partial}{\partial x}$ and with coefficients in~$\K(x)$ form a
non-commutative ring, denoted $\K(\x)\langle \Dx\rangle$, that shares many
algebraic properties with the commutative ring~$\K[\x]$. The structural
analogy between polynomials and linear differential equations was discovered
long ago by Libri and Brassinne~\cite{Libri1833, Brassinne1864, Demidov83}.
They introduced the bases of a non-commutative elimination theory, by defining
the notions of greatest common right divisor (GCRD) and least common left
multiple (LCLM) for differential operators, and by designing a Euclidean-type
algorithm for GCRDs.
This was  formalised by Ore~\cite{Ore32,Ore33}, who set up a common
algebraic framework for polynomials and linear differential operators (and
other skew polynomials, including difference and $q$-difference operators).
Yet, the algorithmic study of linear differential operators is currently much
less advanced than in the polynomial case. The cost of product in
$\K(\x)\langle \Dx\rangle$ has been addressed only recently
in~\cite{VdHoeven02,BoChLe08}. 

The general aim of this work is to take a step towards a systematic study of
the complexity of operations in $\K(\x)\langle \Dx\rangle$. We promote the
idea that (polynomial) matrix multiplication may well become the common
yardstick for measuring complexities in this non-commutative setting. The
specific goal of the present article is to illustrate this idea
for LCLMs. We focus on LCLMs since several higher level
algorithms rely crucially on the efficiency of this basic computational
primitive. For instance, algorithms for manipulating D-finite functions
represented by annihilating equations use common left multiples for performing
addition~\cite{Stanley80,SaZi94}. 
LCLMs of several operators are also needed as a basic task in various other higher-level algorithms~\cite{BaChLo03,Le03,ClHo04}. 
Our approach is based on using complexity analysis as a tool for algorithmic
design, and on producing tight size bounds on the various objects involved in
the algorithms.

It is folklore that Ore's non-commutative Euclidean algorithm is
computationally expensive; various other algorithms for computing common left
multiples of two operators have been
proposed~\cite{Heffter1896,Stanley80,SaZi94,Li98,Bostan03,AbLeLi05,VdHoeven11}.
As opposed to Ore's algorithm, these alternatives have the common
feature that they reduce the problem of computing LCLMs to linear algebra.
However, few complexity analyses~\cite{Giesbrecht92,Giesbrecht98,Bostan03,VdHoeven11} and performance comparisons~\cite{Li98,AbLeLi05} are available.

\begin{figure*} \begin{center} \renewcommand{\arraystretch}{1.2}
\tabcolsep4pt
	\begin{tabular}{|c|ccccc|} \hline
{Algorithm}  & {\, \bf Heffter's + DAC$^\star$ \, } & {\,\bf Li's + DAC$^\star$ \,} & {\,\bf van der Hoeven's + DAC\,} & {\,\bf van Hoeij's\,} & {\,\bf {\red New}$^\star$\,} \\
{Complexity} & 
$\widetilde{O}\left( k^{5} r^{4} d\right)$ & 
$\widetilde{O}\left( k^{\theta+3} r^{\theta+2} d\right)$ &
$\widetilde{O}\left( k^{5} r^{4} d\right)$ & 
$\widetilde{O}\left( k^{\theta+1} r^{\theta+1} d\right)$ &
$\widetilde{O}\left( k^{2\theta} r^\theta  d\right)$ \\ \hline
\end{tabular}
\caption{Costs of various algorithms for the LCLM computation of~$k$ operators of bidegrees $(d,r)$ in $(x,\partial)$.
Algorithms marked by a star ($^\star$) also compute cofactors for the same complexity.
}\label{fig:complexity}
\end{center}
\vskip-15pt
\end{figure*}

\medskip \noindent {\bf Main contributions.} As a first contribution, we
design a new algorithm for computing the LCLM of \emph{several\/} operators.
It reduces the LCLM computation to a linear algebra problem on a polynomial
matrix. The new algorithm can be viewed as an adaptation of Heffter's
algorithm~\cite{Heffter1896} to several operators. At the same time, we use
modern linear-algebra algorithms~\cite{Storjohann03,StVi05} to achieve a low
arithmetic complexity. Our algorithm is similar in spirit to Grigoriev's
method~\cite[\S5]{Grigoriev90} for computing GCRDs of several operators.

Before stating more precisely our main results, we need the following
conventions and notations, that will be used throughout the paper. All
algorithms take as input linear differential operators \emph{with
polynomial coefficients}, that is, belonging to $\K[x]\langle \Dx\rangle$,
instead of rational function coefficients. From the computational viewpoint,
this is not too severe a restriction, since the rational coefficients case 
is easily reduced to that of polynomial coefficients, by normalisation. For
$L_1,\ldots,L_k$ in $\K[x]\langle \Dx\rangle$, we write
$\lclm(L_1,\ldots,L_k)$ 
for the primitive LCLM of $L_1,\ldots,L_k$ in  $\K[\x]\langle \Dx\rangle$.
We say that an operator $L\in \K[x]\langle \Dx\rangle$ has bidegree at most
$(d,r)$ in $(x,\partial)$ if it has order at most~$r$ and polynomial
coefficients of degree at most~$d$.
The cost of our algorithms is measured by the number of
arithmetic operations that they use in the base field~$\K$. The constant
$\theta \in [2,3]$ stands for a feasible exponent for matrix multiplication
over~$\K$ (see definition in~Section~\ref{SECT:pre}), and the soft-O notation $\bigOsoft(\,)$ indicates that
polylogarithmic factors are neglected. 
Our main result is the following.
\begin{theorem} \label{th:main}
	Let $L_1, \ldots, L_k$ be  operators in $\K[x]\langle
\Dx\rangle$
of bidegrees at most $(d,r)$ in $(x,\partial)$.
Then~$\lclm(L_1,\ldots,L_k)$ has order at most~$kr$, degrees in
$x$ at most~$dk(rk-r+1)$, and it can be computed in $\bigOsoft(k^{2\theta}
r^\theta d)$ arithmetic operations in~$\K$. 
\end{theorem}

The upper bound $dk(rk-r+1)$ on coefficient degrees is sharp, in the sense that it is reached on \emph{generic\/} inputs. 
It improves by one order of magnitude the best bound $O(k^2r^2d)$ obtained using
previous algorithms. Moreover, for fixed $k$, the cost of the new
algorithm is almost optimal, in the sense that it nearly matches the
arithmetic size of the LCLM. 

As a second contribution, we analyse the worst-case arithmetic complexity of
existing algorithms for LCLMs, as well as the size of their outputs. For
instance, we show that the extension to several operators 
of the ``folklore'' algorithm
in~\cite{Stanley80,SaZi94}  has complexity
$\bigOsoft(k^{\theta+1}r^{\theta+1}d)$. We call this extension \emph{van
Hoeij's algorithm}, after the name of the implementor in Maple's package
\verb+DEtools+ of one of its variants. These estimates are in accordance with our experiments showing
that our new algorithm performs faster for large order~$r$, while van Hoeij's
algorithm is well suited for large~$k$. 

Using our tight degree bounds, we also show that any algorithm that
computes the LCLM of two operators of bidegree~$(d,r)$ in $(x,\pa)$ in
complexity $\bigOsoft(r^\alpha d^\beta)$ can be used as the building block of
a divide-and-conquer (DAC) algorithm that computes the LCLM of $k$ operators
of bidegree~$(d,r)$ in complexity $\bigOsoft(k^{\alpha+2\beta}
r^{\alpha+\beta} d^\beta)$. The costs of several algorithms
are summarised in Figure~\ref{fig:complexity}, where notation {\bf
$\mathcal{A}$ + DAC} indicates that algorithm {\bf $\mathcal{A}$} is used in
a DAC scheme.

As a third contribution, we prove an upper bound $B \approx 2k(d+r)$ on the total degree in $(x,\partial)$ of nonzero common left multiples
(not necessarily of minimal order). 
This is a new instance of the philosophy, initiated
in~\cite{BoChLeSaSc07}, of relaxing order minimality for linear differential
operators, in order to achieve better arithmetic size. While, by
Theorem~\ref{th:main}, the total arithmetic size of the LCLM is typically
$k^3r^2d$, there exist common left multiples of total size $4k^2 (d+r)^2$
only.

A fourth contribution is a fast Magma implementation
that outperforms Magma's  LCLM routine.
Experimental results 
confirm that the practical complexity of the new algorithm behaves as predicted by our
theoretical results. 

Last, but not least, we have undertaken an extensive bibliographic search,
which we now proceed to describe.

\medskip \noindent {\bf Previous work.} Libri~\cite{Libri1833} and
Brassinne~\cite{Brassinne1864} (see also~\cite{Demidov83}) defined the notions
of GCRD and LCLM of linear differential operators, and sketched a
Euclidean-type algorithm for GCRDs. Von
Escherich~\cite{Escherich1883} defined the related notion of differential
resultant of two linear differential operators. 
Articles~\cite{Brassinne1864,Escherich1883} contain the embryo of an algorithm for the LCLM
based on linear algebra; that algorithm was explicitly stated by
Heffter~\cite{Heffter1896}, and later rediscovered by Poole in his classical
book~\cite{Poole36}. The roots of a subresultant theory for differential
operators are in Pierce's articles~\cite{Pierce1903,Pierce1904}.
Blumberg~\cite{Blumberg1912} gave one of the first systematic accounts of the
algebraic properties of linear differential operators. Building on
predecessors' works, Ore~\cite{Ore32,Ore33} extended the Euclidean-type theory
to the more general framework of \emph{skew polynomials}. He
showed~\cite[Theorem~8, \S3]{Ore33} that, while the LCLM is not related to the
GCRD by a simple formula as in the commutative case, there nevertheless exists
a formula expressing the LCLM in terms of the successive remainders in the
Euclidean scheme.
Almost simultaneously, Wedderburn~\cite[\S7-8]{Wedderburn1932} showed that the LCLM can also be computed
by an \emph{extended\/} version of the Euclidean-Ore algorithm, that computes
Bézout cofactors along the way.

In the computer algebra literature, algorithmic issues for skew polynomials
emerged in the 1990s, and were popularised by Bronstein and
Petkov{\v{s}}ek~\cite{BrPe94,BrPe96}. 
Grigoriev~\cite[\S6]{Grigoriev90} designed a fast algorithm for computing the
GCRD of a family of linear differential operators; to do so, he proved tight
bounds on the degree of the GCRD, by extending von Escherich's construction of
the Sylvester matrix for two differential operators to an arbitrary number of
operators. The bound is linear in the number of operators, in their maximal
order and in their maximal degree~\cite[Lemma~5.1]{Grigoriev90}.
Giesbrecht analysed the complexity of the
LCLM computation for two operators, but only in terms of their order~\cite{Giesbrecht92,Giesbrecht98}.
(Strictly speaking, his method was proposed for a different Ore ring, but it
extends to more general settings, including the differential case.) For two
operators $L_1, L_2\in \K(\x)\langle \Dx\rangle$ of orders at most~$r$, the first (Heffter-style)
algorithm~\cite[Lemma~5]{Giesbrecht92}  computes $\lclm(L_1,L_2)$ in
$O(r^\theta)$ operations in~$\K(x)$, while the second
one~\cite[Lemma~2.1]{Giesbrecht98} (based on the extended Euclidean-Ore scheme) uses
$\widetilde{O}(r^2)$ operations in~$\K(x)$. To our knowledge, no algorithm currently exists similar
to the Lehmer-Knuth-Sch\"onhage half-gcd algorithm~\cite[Chapter~11]{GaGe03},
using a number of operations in $\K(x)$ that is quasi-linear in~$r$. 
Li~\cite{Li98} pointed out that algorithms for the LCLM computation that have
good complexity with respect to the order, such as the naive Euclidean-Ore
algorithm, do not necessarily behave well because of coefficient growth.
He developed a generalisation of the classical subresultant theory to Ore
polynomials, that provides determinantal formulas and degree bounds for the
GCRD and the LCLM~\cite{Li98}. He also compared the practical
efficiency of Maple implementations of several algorithms.

Giesbrecht and Zhang~\cite[Theorem~2.1]{GiZh03} mention a complexity
bound of $O(r^5 d^2)$ for the LCLM computation of two operators of bidegree
$(d,r)$ in $(x,\Dx)$, based on an unpublished 2002 note of Li.
Over fields of characteristic zero,
Bostan~\cite[Chapter~10]{Bostan03} sketched a general strategy for computing
several constructions on differential operators (including LCLMs), based on an
evaluation-interpolation approach on power series solutions. He stated,
without proofs, several degree bounds and complexity results. For two
operators $L_1,L_2$ of bidegree $(n,n)$ in $(x,\Dx)$, he announced that using
fast Hermite-Padé approximation for the interpolation step yields an algorithm
that computes $\lclm(L_1,L_2)$ in $O(n^{\theta+2})$ operations. 
The approach was enhanced by van der Hoeven~\cite{VdHoeven11},
who showed that the costs of the basic operations on
differential operators can be expressed in terms of the cost of multiplication
in $\K[\x]\langle \Dx\rangle$, and proved
the complexity bound $O(n^{\theta+2})$ stated without proof in~\cite[\S10.5]{Bostan03}.

\section{Preliminaries} \label{SECT:pre} Let $(K,\delta)$ be a differential
field, that is, a field $K$ equipped with an additive map $\delta : K
\rightarrow K$ satisfying the Leibniz rule $\delta(xy) = x\delta(y) +
\delta(x)y$ for all $x,y\in K$. We denote by~$K[\pa ; \delta]$ the ring of
linear differential operators over the differential field~$(K, \delta)$. 
A nonzero element~$L$ in~$K[\pa ; \delta]$ is of the form \[ L = a_r
\pa^r + a_{r-1} \pa^{r-1} + \cdots + a_0, \] where~$a_r, a_{r-1}, \ldots, a_0
\in K$ with~$a_r \neq 0$. We call~$r$ the \emph{order} of~$L$, and denote it
by~$\ord(L)$.
The
noncommutative ring~$K[\pa ; \delta]$ is a left (and right) principal ideal
domain, for which a Euclidean algorithm exists~\cite{Ore32,Ore33}.

Let~$L, L_1, \ldots, L_k$ be nonzero elements in~$K[\pa; \delta]$. Then~$L$ is
called a \emph{common left multiple\/} (CLM) of~$L_1, \ldots, L_k$ if~$L = Q_1 L_1 =
\cdots = Q_k L_k$ for some~$Q_1, \ldots, Q_k \in K[\pa; \delta]$. A common
left multiple of the least order is called a \emph{least common left multiple\/} (LCLM). 
Two LCLMs of~$L_1$, \ldots,~$L_k$ are $K$-linearly
dependent.

\smallskip
Our main focus is on the particular case $K = \K(x)$, the field of
rational functions with coefficients in $\K$, and $\delta = \frac{d}{dx}$, the
usual derivation with respect to~$x$. In this case, we use the notation
$\K(\x)\langle \Dx\rangle$ for $K[\pa ; \delta]$, and $\lclm(L_1,\ldots,L_k)$
for the primitive LCLM of $L_1,\ldots,L_k$ in  $\K[\x]\langle \Dx\rangle$,
that is the LCLM of $L_1,\ldots,L_k$ computed in $\K(\x)\langle \Dx\rangle$ and normalised in $\K[\x]\langle \Dx\rangle$ with trivial content.
However, in order to keep the mathematical exposition as independent as
possible of any particular case, we stick to the more general setting $K[\pa;
\delta]$ whenever we discuss mathematical properties and bird's-eye view
descriptions of algorithms.

\paragraph*{\bf Polynomial and matrix arithmetic} The cost of our algorithms
will be measured by the number of field operations in $\K$ they use. To
simplify the presentation, we assume that polynomials in $\K[x]_{<n}$ (\ie, of degree
less than~$n$ in~$x$) can be multiplied within $O(n \log (n)\,\log\log (n))= \bigOsoft
(n)$ operations in~$\K$, using the FFT-based algorithms
in~\cite{ScSt71,CaKa91}. Most basic polynomial operations in $\K[x]_{<n}$
(division, extended gcd, interpolation, etc.) have cost
$\bigOsoft(n)$~\cite{GaGe03}. We suppose that $\theta$ is a feasible exponent
for matrix multiplication over~$\K$, that is, a real constant $2 \leq \theta 
\le 3$, such that two $n\times n$ matrices with coefficients in $\K$ can be
multiplied in time $O(n^\theta)$. The current tightest upper bound is
$\theta < 2.3727$~\cite{VassilevskaWilliams11}, following work of
Coppersmith and Winograd~\cite{CoWi90}, and Stothers~\cite{Stothers10}.

The following result, due to Storjohann and
Villard~\cite{Storjohann03,StVi05}, will be helpful to estimate complexities
for solving linear systems arising from LCLM computations. Note that this is
currently the best complexity result on polynomial linear algebra. 
The probabilistic aspects of the algorithms described
in this article are entirely inherited from it. 

\begin{theorem}\emph{\cite{Storjohann03,StVi05}} \label{theo:SV}
Let~$M$ be an $m \times n$
matrix with entries in $\K[x]_{< d}$. 
The rank~$\rho$ of~$M$ can be computed together with~$m-\rho$
linearly independent polynomial elements in the left kernel of~$M$ within
$ \widetilde{O} ( {mn}\,  {\rho^{\theta-2}} \, d )$ operations in $\K$
by a (certified) randomised Las Vegas algorithm.

Moreover, if $m=n$, then the determinant of~$M$ can be computed using $\bigOsoft(n^\theta \, d)$ operations in $\K$.
\end{theorem}

\section{Linear formulation for \\ common left multiples} \label{SECT:mat}

In order to connect the computation of common left multiples with linear algebra, we introduce some more notation.
For a nonnegative integer~$n$, we denote by~$K[\pa;\delta]_{\le n}$ the $K$-linear
subspace of~$K[\pa; \delta]$ consisting of all linear differential operators whose orders are at most~$n$.
Moreover, we define a $K$-linear bijection
$$\begin{array}{cccc}
\phi_n: & K[\pa;\delta]_{\le n} & \longrightarrow & K^{n+1} \\ \\
     & \sum_{i=0}^n a_i \pa^i & \mapsto & (a_n, a_{n-1}, \ldots, a_0).
\end{array}
$$
For a nonzero element~$P \in K[\pa; \delta]$ of order~$m$, and for an integer~$n$ with~$n\ge m$, we define
the Sylvester-type matrix
\[ S_n(P) := \left( \begin{array}{c}
 \phi_n\left(\pa^{n-m} P \right) \\
 \phi_n\left(\pa^{n-1-m} P \right) \\
 \vdots \\
 \phi_n(P)
 \end{array} \right). \]
The matrix~$S_n(P)$
has~$n-m+1$ rows and~$n+1$ columns. In particular,~$S_n(1)$ is the identity
matrix of size~$n+1$. This matrix enables one to express multiplication by $P$ in~$K[\pa; \delta]$ as a vector-matrix product. Precisely, for~$Q \in K[\pa; \delta]_{ \le n-m}$,
\begin{equation} \label{EQ:multiply}
 \phi_n(QP) = \phi_{n-m}(Q) S_n(P).
 \end{equation}

Let~$L_1, \ldots, L_k$ be nonzero elements in~$K[\pa; \delta]$.
For $n \ge \max_{1 \le i \le k} \ord(L_i)$, the  matrix
\begin{equation} \label{EQ:mat}
 M_n := \left(
\begin{array}{cccc}
S_n(L_1) &    &  &   \\
         & S_n(L_2) &  & \\
         &  &  \ddots &  \\
         &  &         & S_n(L_k) \\
S_{n}(-1) & S_{n}(-1) & \cdots & S_{n}(-1)
\end{array}\right)
\end{equation}
has~$(k+1)(n+1)- \sum_{i=1}^k \ord(L_i)$ rows and~$k(n+1)$ columns.

\smallskip The following theorem is the main result of this section.
\begin{theorem} \label{TH:mat}
Let~$L_1, \ldots, L_k$ be  elements in~$K[\pa; \delta] \setminus \{ 0\}$ of  orders~$r_1,  \ldots, r_k$,
and let~$n \geq \ord(\lclm(L_1, \ldots, L_k)).$
\begin{enumerate}
\item[(i)]
If~$L$ is a common left multiple of~$L_1, \ldots, L_k$ such that $\ord(L) \le n$ and 
$L=Q_1L_1=\cdots=Q_kL_k,$
then the vector~$\left(\phi_{n-r_1}(Q_1), \ldots, \phi_{n-r_k}(Q_k), \phi_n(L) \right)$
belongs to the left kernel of the matrix~$M_n$ defined in~\eqref{EQ:mat}.
\item[(ii)] If the vector~$(\vu_1, \ldots, \vu_k, \vu)$ is a nonzero vector in the left kernel of~$M_n$,
where~$\vu_i \in K^{n+1-r_i}$ for~$i=1, \ldots, k$ and~$\vu \in K^{n+1}$, then~$\vu$ is nonzero,
and~$\phi_n^{-1}(\vu)$ is a common
left multiple of~$L_1$, \ldots,~$L_k$ with respective left cofactors~$\phi_{n-r_1}^{-1}(\vu_1)$,
\ldots,~$\phi_{n-r_k}^{-1}(\vu_k)$.
\item[(iii)] If~$\rho$ is the rank of~$M_n$, then\\
$ \ord \left( \lclm(L_1, \ldots, L_k) \right) = \rho   + \sum_{i=1}^k r_i - k(n+1). $
\end{enumerate}
\end{theorem}
\myproof Suppose that $L=Q_iL_i$ for~$1 \le i \le k$.
By~\eqref{EQ:multiply},
\[ \phi_{n-r_i}(Q_i) S_n(L_i) = \phi_n(L), \]
which is equivalent to
$ \phi_{n-r_i}(Q_i) S_n(L_i) + \phi_n(L) S_n(-1) = 0. $
Therefore the vector~$(\phi_{n-r_1}(Q_1), \ldots, \phi_{n-r_k}(Q_k), \phi_n(L))$ belongs to the left kernel of~$M_n$.
The first assertion is proved.

Conversely, suppose that~$(\vu_1, \ldots, \vu_k, \vu)$ is a nonzero vector in the left kernel of~$M_n$.
Then~$\vu_i S_n\left( L_i \right) + \vu S_n(-1)=0$ for all~$i$ with~$1 \le i \le k$.
It follows from~\eqref{EQ:multiply} that
$$\phi_{n-r_i}^{-1}(\vu_i) L_i = \phi_n^{-1}(\vu).$$
Thus, $\vu$ is nonzero, for otherwise, since $\phi_n$ is an isomorphism, all the $\vu_i$ would be equal to zero.
The second assertion follows.

To prove the last assertion, we set~$L=\lclm(L_1, \ldots, L_k)$ and~$\ell=\ord(L)$.
Assume further that~$L=Q_i L_i$ for all~$i$ with~$1 \le i \le k$.
Then~$L, \pa L$, \ldots, $\pa^{n-\ell} L$ are common left  multiples of~$L_1,$ \ldots, $L_k$ of orders at most~$n$, and such that
\[  \pa^j L = \left(\pa^j Q_i \right) L_i \quad
\mbox{for all $1 \le i \le k$ and~$0 \le j \le n-\ell$.} \]
By the first assertion, for $0 \leq j \leq n-\ell$, the vector
\[ \vv_j = \left(\phi_{n-r_1}\left( \pa^j Q_1 \right), \ldots, \phi_{n-r_k} \left(\pa^j Q_k \right),
\phi_n\left(\pa^{j} L\right)\right) \]
belongs to the left kernel of~$M_n$. These vectors are $K$-linearly independent because~$L,$ $\pa L$, \ldots, $\pa^{n-\ell} L$
are. 
On the other hand, if~$(\vu_1, \ldots, \vu_k, \vu)$ is a nonzero vector in the left kernel of~$M_n$,
where~$\vu_1 \in K^{n-r_1+1}, \ldots, \vu_k \in K^{n-r_k+1},$ and~$\vu \in K^{n+1}$, then~$\phi_n^{-1}(\vu)$ is
a common left multiple of~$L_1$, \ldots,~$L_k$ with order no greater than~$n$ by the second assertion.
Hence,~$\phi_n^{-1}(\vu)$ is a $K$-linear combination of~$\pa^{n-\ell}L$, $\pa^{n-\ell-1} L$, \ldots,~$L$,
because it is a left multiple of~$L$.
Hence, there exist~$c_{n-\ell}$, $c_{n-\ell-1}$, \ldots, $c_0$ in~$K$ such
that
$$\phi_n^{-1}(\vu) =c_{n-\ell} \pa^{n-\ell} L + c_{n-\ell-1} \pa^{n-\ell-1} L + \cdots + c_0 L,$$
which implies that the last $n+1$ coordinates of the vector
$(\vu_1, \ldots, \vu_k, \vu)-\sum_{j=0}^{n-\ell}c_j \vv_j$
are all equal to zero. Since this vector belongs to
the left kernel of~$M_n$, all its coordinates are zero by the second assertion.
We conclude that $\{  \vv_0,\ldots,  \vv_{n-\ell}\}$ is a $K$-basis of the left kernel of $M_n$, and thus~$n-\ell+1$ is its dimension. Then (iii) follows from the rank-nullity theorem, because~$M_n$
has~$(k+1)(n+1)- \sum_{i=1}^k r_i$ rows.
\foorp

\medskip
Since the rank of~$M_n$ is at most~$k(n+1)$, a direct consequence of Theorem~\ref{TH:mat}~(iii) is the following classical result.
\begin{cor} \label{COR:bounds} 
For~$L_1, \ldots, L_k \in K[\pa; \delta] \setminus \{0\}$,
\[ \ord \left( \lclm(L_1, \ldots, L_k) \right) \le \ord(L_1) + \cdots + \ord(L_k). \]
\end{cor}


\section{Computing LCLMs} \label{SECT:comp}
In this section, we review a few known methods for computing LCLMs and present a new one based on
Theorem~\ref{TH:mat}.

\subsection{Computing an LCLM of two operators} \label{SUBSECT:comp2}
Given two nonzero elements~$L_1$ and~$L_2$ of respective orders~$r_1$ and~$r_2$, we consider various methods for computing their LCLMs.
The first methods  compute left cofactor(s) of
the given operator(s) first, and find an LCLM by multiplication in~$K[\pa; \delta]$. The last method is specific to $K = \K(x)$.

\subsubsection{Heffter's algorithm} 
The first method can be traced back to Brassinne~\cite{Brassinne1864}, von Escherich~\cite{Escherich1883} and Heffter~\cite{Heffter1896}.
The sequence:
\[ \pa^{r_2} L_1, \ldots, \pa L_1, L_1, \, \pa^{r_1} L_2, \ldots, \pa L_2, L_2 \]
has~$r_1 + r_2 + 2$ elements, each of which is of order at most~$r_1+r_2$.
Thus, these elements are $K$-linearly dependent. To compute $\lclm(L_1, L_2)$, the strategy is to 
find the maximal integer~$m$ and corresponding elements~$a_{1,0}, \ldots, a_{1,r_2-m}$, $a_{2,0}, \ldots, a_{2,r_1-m} \in K$
with~$a_{1, r_2-m} a_{2, r_1-m} \neq 0$ such that
\[ \sum_{i=m}^{r_2} a_{1, r_2 -i} \pa^{r_2-i} L_1  + \sum_{j=m}^{r_1} a_{2, r_1-j} \pa^{r_1-j} L_2 = 0. \]
Set
$A_1 = \sum_{i=m}^{r_2} a_{1, r_2 -i} \pa^{r_2-i} \, \text{and} \,
A_2 =  \sum_{j=m}^{r_1} a_{2, r_1-j} \pa^{r_1-j}.$
Then~$A_1 L_1 + A_2 L_2 = 0$. Therefore, the product~$A_1L_1$ is an LCLM of~$L_1$ and~$L_2$ due to the maximality of~$m$.

This method can be reformulated using the notation introduced in Section~\ref{SECT:mat}.
For a vector~$\vv \neq 0$ represented by
\[ (\underbrace{0, \ldots, 0}_k, v_{k+1}, \ldots ), \quad \mbox{where $v_{k+1} \neq 0$,} \]
in a finite-dimensional $K$-vector space equipped with the standard basis
$(1, 0, \ldots, 0), (0,1, 0, \ldots, 0), \ldots, (0, \ldots, 0, 1),$
\sloppy we define~$\cN(\vv)$ to be~$k$.
For~$n \ge \max(r_1, r_2)$, define
\begin{equation} \label{EQ:mat2}
U_{n} = \left( \begin{array}{c}
           S_{n}(L_1) \\
           S_{n}(L_2)
           \end{array} \right).
\end{equation}
Then, Heffter's method consists in computing a vector~$\vv \neq 0$ in the left kernel of~$U_{r_1+r_2}$
such that~$\cN(\vv)$ is maximal.

The next lemma connects the order of~$L=\lclm(L_1, L_2)$ with the rank of~$U_{r_1+r_2}$. 
It easily follows from the observation that a maximal subset of
$K$-linearly independent elements in~$\{ \partial^{r_2} L_1, \ldots, L_1, \partial^{r_1} L_2, \ldots, L_2\}$
consists of~$\partial^{r_2}L_1,$ \ldots, $L_1$ and~$\partial^{\ell - r_2-1} L_2, \ldots, L_2$, where~$\ell = \ord(L)$.
\begin{lemma} \label{LM:prank}
Let~$L_1, L_2$ be two nonzero elements in $K[\pa; \delta]$ of orders~$r_1,r_2$.
Then $\ord \left( \lclm(L_1, L_2) \right) = \rank(U_{r_1+r_2}) - 1.$
\end{lemma}

\subsubsection{Euclidean algorithms}
The second family of methods is based on the Euclidean-Ore algorithm for differential operators~\cite{Ore33}. 

\medskip \noindent {\bf Ore's algorithm.} Assume that~$r_1 \ge r_2$. Setting~$R_1=L_1,$ $R_2=L_2$, one can compute the Euclidean (right) divisions 
\[   R_i = R_{i-2} - Q_i R_{i-1}, \]
for quotients~$Q_i \in K[\pa; \delta]$, and remainders~$R_i \neq 0$ satisfying $\ord(R_i) < \ord(R_{i-1})$ for~$i=3, \ldots, m$, and $R_{m+1}=0$. Then, as in the commutative case, $R_m$ is shown to be the GCRD of $L_1$ and $L_2$. 
Ore~\cite[\S2]{Ore33} proved that the following product 
\begin{equation} \label{eq:Ore} R_{m-1} R_m^{-1}
R_{m-2} R_{m-1}^{-1} \cdots R_3 R_4^{-1} R_2 R_3^{-1} R_1 
\end{equation} is an LCLM of $L_1$ and $L_2$. (Here $AB^{-1}$ denotes the exact left quotient of $A$ and $B$, that is $Q$ such that $A=QB$.)

\medskip \noindent {\bf Extended Euclidean-Ore algorithm.} Wedderburn~\cite[\S7-8]{Wedderburn1932} observed (see also~\cite{BrPe94})
that the
computation of~\eqref{eq:Ore} can be avoided, if one replaces the Euclidean
algorithm by its extended version. Precisely, letting $C_1=1$, $C_2=0$, and
 \[C_i=C_{i-2} - Q_i C_{i-1}, \quad \textrm{for} \quad i=3, \ldots,
m,\]  the product~$(C_{m-1} - Q_{m-1} C_m)R_1$ is an LCLM of~$L_1$ and~$L_2$.

\medskip \noindent {\bf Li's determinantal expression.}
As in the commutative case, a more efficient version of the extended Euclidean-Ore algorithm is based on subresultants~\cite[\S5]{Li98}.
To avoid technicalities, we present an alternative, efficient, variant of the subresultant algorithm, 
based on a determinantal formulation~\cite[Proposition~6.1]{Li98}. This method assumes that the order~$g$ of the GCRD of $L_1$ and $L_2$ is already known. Then, one constructs a square matrix $\mathcal{L}$ of size $r_1+r_2-2g+2$ whose first $r_1+r_2-2g+1$ columns are the first $r_1+r_2-2g+1$ columns of the matrix 
$ \left( \begin{array}{c}
           S_{r_1+r_2-g}(L_1) \\
           S_{r_1+r_2-g}(L_2)
           \end{array} \right),
$
and whose last column is the transpose of the vector
$ (\pa^{r_2-g}, \ldots, \pa, 1, \underbrace{0, 0, \ldots, 0}_{r_1-g+1}).$

\noindent If $\det(\mathcal{L})$ is denoted~$U$, then  $UL_1$ is an LCLM of $L_1$ and ~$L_2$.

\subsubsection{Van der Hoeven's algorithm}
The algorithm that we very briefly mention now is specific to the case $K =
\K(x)$, where the base field $\K$ has characteristic zero. It works by evaluation-interpolation. The idea, originating
from~\cite{Bostan03}, is to perform operations on differential operators by
working on their fundamental systems of solutions. Due to space limitations, and in view of its complexity analysis,
we do not give more details here, and refer the reader to the
article~\cite{VdHoeven11}.

\subsection{Computing an LCLM of several operators} \label{SUBSECT:comps}
Given several nonzero operators~$L_1, L_2, \ldots, L_k \in K[\pa; \delta]$ 
we describe various ways to compute~$\lclm(L_1, \ldots, L_k)$.

\subsubsection{Iterative LCLMs}
An obvious method is to compute an LCLM of $k$ operators iteratively, that is,
\begin{equation}\label{eq:iterative}
	L = \lclm \left(L_1, \lclm(L_2, \ldots, \lclm(L_{k-1}, L_k) \right). 
\end{equation}	

A computationally more efficient (though mathematically equivalent) method is by a divide-and-conquer algorithm, based on the repeated use of the formula
\begin{equation} \label{eq:iterativeByDAC}\begin{split}
L =  \lclm \big( & \lclm(L_1, \ldots, L_{\lfloor k/2 \rfloor}), \\	
& \qquad \lclm(L_{\lfloor k/2 \rfloor + 1}, \ldots, L_k) \big).
\end{split}\end{equation}

Of course, the efficiency of an iterative algorithm depends on that of the algorithm used for the LCLM of two operators. This is quantified precisely in Section~\ref{sec:ABC}.

\subsubsection{Van Hoeij's algorithm}
Another algorithm for computing the LCLM of $k$~linear differential operators was implemented by van Hoeij as Maple's \verb+DEtools[LCLM]+ command; it seemingly was never published.
For $k=2$, the method is folklore; it is implicit, for instance, in the proof of~\cite[Theorem~2.3]{Stanley80}. A variant of it is also implemented  by the \texttt{`diffeq+diffeq`} command in Salvy and Zimmermann's \verb+gfun+ package for Maple~\cite{SaZi94}.

Informally speaking, the method consists in considering a generic solution~$h_j$ of~$L_j$ for~$1\leq j\leq k$, then in finding the first linear dependency between the row vectors $\partial^i(h_1,\dots,h_k)=(\partial^i\cdot h_1,\dots,\partial^i\cdot h_k)$.
In order to perform actual computations, these vectors are represented by the canonical forms
$ \left( \rem(\partial^i,L_1),\dots,\rem(\partial^i,L_k) \right), \, \textrm{for } \, i=0,1,\ldots,$
where $\rem(A,B)$~denotes the remainder of the right Euclidean division of~$A$ by~$B$.
Let \[ L = a_n \pa^{n} + a_{n-1} \pa^{n-1} + \cdots + a_0, \]
where~$a_0, a_1, \ldots, a_n$ are undetermined coefficients in $K$.
For all~$i$ with~$1 \le i \le k$, let~$R_i$ be the right remainder
of~$L$ in the division by~$L_i$. Then $L \equiv R_i \mod L_i.$
Since~$L$ has generic coefficients,  $\ord(R_i)$ is equal to $r_i-1$, e.g., by ~\cite[Lemma~2.3]{Li98}.
Note that~$a_0, a_1, \ldots, a_n$ depend linearly on the coefficients of the~$R_i$'s.
There are $s=r_1+ \cdots + r_k$ coefficients in~$R_1,\ldots, R_k$.
Equating~$R_i=0$, for $i=1, \ldots, k,$  we obtain a linear system
\[ (a_n, a_{n-1}, \ldots, a_0) H_n = (0, 0, \ldots, 0), \]
where~$H_n$ is an $(n+1) \times s$ matrix over~$K$. Thus, computing~$\lclm(L_1, \ldots, L_k)$
amounts to computing a nontrivial vector~$\vv$ in the left kernel of~$H_s$ with~$\cN(\vv)$ being maximal.
The rank of~$H_s$ is equal to the order of~$\lclm(L_1, \ldots, L_k)$,
e.g., by~\cite[Proposition~4.3]{AbLeLi05}. Note that the original version of van Hoeij's algorithm does not make use of this last fact, and potentially needs to solve more linear systems, thus being less efficient when the LCLM is not of maximal order. 

\subsubsection{The new algorithm}
As a straightforward consequence of Theorem~\ref{TH:mat}, the $\lclm(L_1, \ldots, L_k)$ can be computed
by determining a nontrivial vector~$\vv$ in the left kernel of~$M_s$ given in~equation~\eqref{EQ:mat},
with~$\cN(\vv)$ being maximal. This method computes not only the LCLM, but also its left cofactors~$Q_1$,
$Q_2$, \ldots, $Q_k$, while van Hoeij's algorithm does not
compute any cofactor.

\section{\!\!\!\!\!\! Algorithms, bounds, complexity} \label{sec:ABC}

\begin{figure*} 
\begin{center} \renewcommand{\arraystretch}{3.7}
\tabcolsep2pt
	\begin{tabular}{|c|c|c|} \hline
	{
			\begin{minipage}{5.3cm}
			\smallskip
			\center{\bf Heffter's algorithm} 

			\medskip
			\begin{enumerate}
			\item Compute the matrix~$U_{r_1+r_2}$ defined in~\eqref{EQ:mat2}. 
			\item Determine its rank~$\rho$; set $\ell := \rho - 1$. \label{step:rankHeffter}
			\item Extract submatrix~$U_\ell$ of~$U_{r_1+r_2}$.
			\item Find the 1-dim kernel $\mathcal{K}$ of~$U_\ell$. \label{step:kernelHeffter} 
			\item Construct 
			$Q_1$ 
			from the first~$\ell-\ord(L_1)+1$ coordinates of $\mathcal{K}$.
			\item Compute and return $Q_1L_1$.\label{step:mulHeffter}
			\end{enumerate}
			\end{minipage}
	}
		& 
		{
		\begin{minipage}{5.3cm}
		\smallskip
		\center{\bf van Hoeij's algorithm} 
		\smallskip
		\begin{enumerate}
		\item For all~$0 \le i \le  s$ and~$1 \le j \le k$, compute
		\begin{align*} c_{i}^{-1} h_{i, j} = \text{rem}(\pa^i, L_j), \; \textrm{where} \\ 
		c_{i} \in \K[x] \; \text{and} \; h_{i,j} \in \K[\x]\langle \Dx\rangle.
		\end{align*}
		\item View the $h_{i,j}$ as rows in $\K[x]^{r_j}$; compute  rank $\rho$ of	$H_s := (h_{i,j})$.\label{step:rankHoeij}
		\item Extract  submatrix~$H_\rho$ of~$H_s$.
		\item Find the 1-dim kernel $\mathcal{K}$ of~$H_\rho$.\label{step:kernelHoeij}
		\item Construct the LCLM from $\mathcal{K}$.
		\end{enumerate}
		\end{minipage}
}	
		&   
		{
		\begin{minipage}{5.3cm}

		\medskip
		\center{\bf Our new algorithm}

		\medskip
		\begin{enumerate}
		\item Compute~$M_s$ defined in~\eqref{EQ:mat}. 
		\item Determine  its rank~$\rho$ ; set~$\ell := \rho + s - k(s+1)$.\label{step:rankNew}%
		\item Extract submatrix~$M_\ell$ of~$M_s$.
		\item Find the 1-dim kernel $\mathcal{K}$ of~$M_\ell$.\label{step:kernelNew}
		\item Construct the LCLM from the last~$\ell+1$ coordinates of $\mathcal{K}$.
		\item Return the LCLM.
		\end{enumerate}
		\end{minipage}
} \\ 
\hline
\end{tabular}
\caption{Pseudo-code for Heffter's algorithm, van Hoeij's algorithm and our new algorithm.}
\label{fig:Heffter+vH+new}
\end{center}
\end{figure*}

In this section, we let~$K = \K(x)$ be the field of rational functions with
coefficients in a field~$\K$, and $\delta = \frac{d}{dx}$ be the usual
derivation with respect to~$x$. Recall that in this case we use the notation
$\K(\x)\langle \Dx\rangle$ for $K[\pa ; \delta]$, and $\lclm(L_1,\ldots,L_k)$
for the \emph{primitive\/} LCLM of $L_1,\ldots,L_k$ in  $\K[\x]\langle \Dx\rangle$.

All algorithms analysed below are specialisations of the algorithms reviewed
in the previous section to $K = \K(x)$. Moreover,  we make the non-restrictive assumption that all algorithms take as input linear differential operators
\emph{with polynomial coefficients}, that is, belonging to $\K[x]\langle
\Dx\rangle$.

The degree of a nonzero operator $L \in \K[\x]\langle \Dx\rangle$, denoted $\deg_x(L)$,
is defined as the maximal degree of its coefficients.
As in the case of usual commutative polynomials,
\[ \deg_x (AB) = \deg_x (A) + \deg_x (B) \; \mbox{for all~$A, B \in \K[\x]\langle \Dx\rangle \setminus \{ 0 \}.$} \]

\subsection{Tight degree bounds for the LCLM}
First, we give a sharp degree bound for LCLMs. As we show later, this bound improves upon the bound that can be derived from van Hoeij's algorithm.
\begin{theorem} \label{TH:db}
Let~$L_1, \!\ldots \!, \! L_k$ be operators in~$\K[\x]\langle \Dx\rangle \setminus \{ 0\}$. Let
$ s= \ord(L_1) + \cdots+ \ord(L_k), \quad \textrm{and} \quad d = \max_{i=1}^k \deg_x (L_i).$
If~$L=\lclm(L_1, \ldots, L_k)$, then
$ \deg_x (L) \le d (k(s+1)-s). $
\end{theorem}

\myproof
By Corollary~\ref{COR:bounds}, $\ord(L) \le s$. It follows from Theorem~\ref{TH:mat}
and Cramer's rule
that every nonzero coefficient of~$L$ is a quotient of two 
minors of~$M_s$. Note that every square submatrix of~$M_s$ has size at
most~$k(s+1)$, since~$M_s$ has~$k(s+1)+1$ rows and~$k(s+1)$ columns. Thus, the
degree of the determinant of such a submatrix is bounded by~$d(k(s+1)-s)$,
because every entry of~$M_s$ is of degree at most~$d$, and the
last~$s+1$ rows of~$M_s$ are free of~$x$. \foorp

As a consequence of Corollary~\ref{COR:bounds} and Theorem~\ref{TH:db}, 
the first part of Theorem~\ref{th:main} is easily deduced.

\subsection{LCLMs of two operators}
The following result encapsulates complexity analyses of LCLM algorithms for two operators. Heffter's, van Hoeij's and our new algorithm are summarised in Figure~\ref{fig:Heffter+vH+new}.

\begin{theorem}\label{theo:compl-2ops}
Let $L_1, L_2 \in \K[\x]\langle
\Dx\rangle$ be operators of bidegrees at most $(d,r)$ in $(x,\Dx)$.	
Then it is possible to compute the LCLM of $L_1$ and $L_2$ in complexity
\begin{tabbing}
	\quad \emph{(a)} $\bigOsoft(\min(r^\theta \, d^2, r^3 \, d))$ by Heffter's and van der Hoeven's \\ algorithms,\\
	\quad \emph{(b)} $\bigOsoft(r^{\theta+1} \, d)$ by Li's and by van Hoeij's algorithms,\\
	\quad \emph{(c)} $\bigOsoft(r^{\theta} \, d)$ by the new algorithm.
\end{tabbing}
\end{theorem}

\myproof By~\cite[Theorems~5,~8 \& 23]{VdHoeven11}, and using bounds from
Theorem~\ref{TH:db}, the complexity of van der Hoeven's algorithm is
$\bigOsoft(\min((rd)^2 \, r^{\theta-2}, r^3d))$.
The most costly parts of Heffter's algorithm are
Steps~\ref{step:rankHeffter},~\ref{step:kernelHeffter}
and~\ref{step:mulHeffter}. Since the matrix~$U_{r_1+r_2}$ has size~$O(r)$ and polynomial coefficients of degree at most~$d$, the rank and kernel computations involved in Steps~\ref{step:rankHeffter} and~\ref{step:kernelHeffter} can be performed using $\bigOsoft(r^\theta d)$ operations, by Theorem~\ref{theo:SV}. Step~\ref{step:mulHeffter} consists in 
multiplying two operators in $\K[\x]\langle \Dx\rangle$ of bidegrees at most $((r-1)d,r)$ and $(d,r)$ in $(x,\partial)$. This can be done using
$\bigOsoft(\min((rd)^2 \, r^{\theta-2}, r^3d))$. This proves (a). 

The dominant parts of Li's algorithm are the computation of $g=\ord(
\text{GCRD}(L_1,L_2))$, and the expansion of $O(r)$ minors of a polynomial matrix of size $O(r)$ and degree at most~$d$. By using~\cite[Lemma~5.1]{Grigoriev90} and Theorem~\ref{theo:SV}, $g$ can be computed using $\bigOsoft(r^\theta d)$ operations in $\K$, and the minors can be expanded in  $\bigOsoft(r^{\theta +1} d)$.
The dominant parts of van Hoeij's algorithm are
Steps~\ref{step:rankHoeij} and~\ref{step:kernelHoeij}.
Since $k=2$, matrix~$H_s$ has size $O(r)$. By an easy induction, its $(r+j)$th row has polynomial coefficients of degrees at most~$2jd$, thus $H_s$ has degree $O(dr)$. By Theorem~\ref{theo:SV}, the rank and kernel computations have complexity $\bigOsoft(r^{\theta+1}d)$. This proves (b).

The dominant parts of the new algorithm are
Steps~\ref{step:rankNew} and~\ref{step:kernelNew}. Since $k=2$, the polynomial matrix~$M_s$ has size $O(r)$ and degree at most $d$. By Theorem~\ref{theo:SV} again, the rank and kernel computations have cost $\bigOsoft(r^{\theta}d)$. This completes the proof.
\foorp

Quite surprisingly, the costs of Heffter's and of van der Hoeven's
algorithms are penalised by the complexity of multiplication of operators,
which is not well-understood yet for general bidegrees. Precisely,
it is an open problem whether two operators of bidegree $(d,r)$ in
$(x,\partial)$ can be multiplied in nearly optimal time $\bigOsoft(
r^{\theta-1} d)$. If such an algorithm were discovered, then the costs of both
algorithms would become $\bigOsoft(r^{\theta} \, d)$, improving the
corresponding entries in Figure~\ref{fig:complexity}.

\subsection{LCLMs of several operators} \label{ssec:cost-sev} 
We analyse three algorithms for LCLMs of several operators: 
DAC, van Hoeij's and our new algorithm. 

\subsubsection{LCLMs by divide-and-conquer} \label{ssec:iter}

\vspace{-0.2cm}

\begin{theorem} \label{theo:DAC}
	Suppose that we are given an algorithm  computing the LCLM
of two differential operators which, on input $L_1,L_2 \in \K[\x]\langle
\Dx\rangle$ of bidegree at most $(D,R)$ in $(x,\Dx)$, computes $\lclm(L_1,L_2)$
in complexity $\bigOsoft (R^\alpha D^\beta)$ for some constants $\alpha \geq
2$ and $\beta \geq 1$ independent of~$D$ and~$R$. 

There exists an algorithm which, on input $L_1,\ldots, L_k \in \K[\x]\langle
\Dx\rangle$ of bidegrees at most $(d,r)$ in $(x,\Dx)$, computes $L = \lclm(L_1,\ldots,L_k)$ using $\bigOsoft (k^{\alpha+2\beta} r^{\alpha+\beta} d^\beta)$ operations in $\K$.
\end{theorem}

\myproof
Suppose without loss of generality that $k=2^\ell$ is a power of~$2$.	
To compute $L$, we  use a strategy based on~\eqref{eq:iterativeByDAC}, similar to that of the subproduct tree~\cite[\S10.1]{GaGe03}: we partition the family $(L_1, \ldots, L_k)$ into pairs, compute the LCLM of each pair using algorithm~$\mathcal{A}$ available for two operators, remove the polynomial content, then compute LCLMs of pairs, and so on. Let $L_{[a:b]}$ denote the LCLM of $L_a, \ldots, L_{b}$ with the content removed.
At level~1, the algorithm computes the $k/2$ operators $L_{[1:2]}, \ldots, L_{[k-1:k]}$, at level~2 the $k/4$ operators $L_{[1:4]}, \ldots, L_{[k-3:k]}$, and so on, the last computation at level~$\ell$ being that of $L$ as the LCLM of $L_{[1:k/2]}$ and $L_{[k/2+1:k]}$. Let $\C(R,D) = \bigOsoft (R^\alpha D^\beta)$ denote the complexity of algorithm~$\mathcal{A}$ on inputs of bidegrees at most $(D,R)$.
By Theorem~\ref{TH:db}, the operators computed at level~$1\leq j\leq \ell$ have bidegree at most~$(2^j r, 2^j d((2^j-1)r+1))$.
Thus, the total cost of the DAC algorithm on $k$ inputs of bidegree at most $(d,r)$ is bounded by
$ \sum_{j=0}^{\ell - 1} \frac{k}{2^{j+1}} \cdot \C \left( 2^j r, 2^jd ((2^j-1)r+1) \right), $
plus the cost of the content removal, which is negligible. 
Up to polylogarithmic factors, the cost is bounded by
$ \displaystyle{\sum_{j=0}^{\ell - 1} \frac{k}{2^{j+1}} \cdot (2^j r)^\alpha \cdot (4^j dr)^\beta = \frac{k}{2} \cdot r^{\alpha + \beta} \cdot d^\beta \cdot \sum_{j=0}^{\ell - 1}  (2^j)^{\alpha+2\beta-1}, }$
which is $O(k^{\alpha+2\beta} r^{\alpha+\beta} d^\beta)$. This concludes the proof.
\foorp	

The cost of the algorithm is essentially that of its last step; this is a typical feature of DAC
algorithms.
A similar analysis shows that the iterative algorithm based on
formula~\eqref{eq:iterative} is less efficient, and has complexity
$\bigOsoft(k^{\alpha+2\beta+1} r^{\alpha+\beta} d^\beta)$.

\smallskip As a corollary of Theorems~\ref{theo:compl-2ops}
and~\ref{theo:DAC}, we get a proof of the complexity estimates in the
first three entries of Figure~\ref{fig:complexity}.

\subsubsection{Van Hoeij's and the new algorithm}
\vspace{-0.3cm}
\begin{theorem}\label{theo:compl-kops}
Let $L_1, \ldots, L_k \in \K[\x]\langle
\Dx\rangle$ have bidegrees at most $(d,r)$ in $(x,\Dx)$.	
One can compute $\lclm(L_1,\ldots,L_k)$ 
\begin{tabbing}
	\quad \emph{(a)} in $\bigOsoft(k^{\theta+1} r^{\theta+1} \, d)$ operations by van Hoeij's algorithm,	\\
	\quad \emph{(b)} in $\bigOsoft(k^{2\theta} r^{\theta} \, d)$ operations by the new algorithm. 
\end{tabbing}
\end{theorem}

\myproof
The proof is similar to that of~Theorem~\ref{theo:compl-2ops}(b) and~(c).
The most costly parts of van Hoeij's algorithm are
Steps~\ref{step:rankHoeij} and~\ref{step:kernelHoeij}.
Matrix~$H_s$ has size $O(kr)$ and  polynomial coefficients of degree~$O(k rd)$. By Theorem~\ref{theo:SV}, the rank and kernel computations have complexity $\bigOsoft((kr)^{\theta} \, krd) = \bigOsoft(k^{\theta+1} r^{\theta+1} d)$. This proves (a).
The dominant parts of the new algorithm are
Steps~\ref{step:rankNew} and~\ref{step:kernelNew}. The polynomial matrix~$M_s$ has size $O(k^2 r)$ and degree at most $d$. By Theorem~\ref{theo:SV}, the rank and kernel computations have cost $\bigOsoft ((k^2 r)^{\theta}d) = \bigOsoft(k^{2\theta} r^\theta d)$. 
\foorp

\smallskip
As a corollary of Theorem~\ref{theo:compl-kops}, we
get a proof of the complexity estimates in the last two entries of Figure~\ref{fig:complexity}. Note that Cramer's rule applied to the matrix $H_s$ analysed in the previous proof yields the bound $O(k^2r^2d)$ on the coefficient degrees of the LCLM. This bound is improved by Theorem~\ref{th:main}.

\vspace{-0.1cm}

\begin{section}{\!\!\!\! Smaller common left multiples}

Our approach to computing more common left multiples (CLMs), that are generally
not of minimal order, but smaller in total arithmetic size than the LCLM, is
similar to the linear-algebraic approach used in Section~\ref{SECT:mat}.
However, instead of considering a matrix encoding the~$\partial^iL_j$, with
\emph{polynomial\/} coefficients, we turn our attention to a matrix encoding
the~$x^{i_1}\partial^{i_2}L_j$, with \emph{constant\/} coefficients.

\medskip \noindent {\bf Existence of smaller CLMs.} The new building block to consider is, for an operator~$P$ in~$\K[\x]\langle
\Dx\rangle$ of total degree~$\Delta$ in $x$ and~$\partial$, and
an integer~$N\geq\Delta$, the $\binom{N-\Delta+2}2\times\binom{N+2}2$
matrix~$C_N(P)$ with scalar coefficients whose rows represent the operators
of the form~$x^{i_1}\partial^{i_2}P$ for~$0\leq i_1+i_2\leq N$, in any fixed
order, and whose columns are indexed by the monomials of total degree at
most~$N$, in any fixed order.

Let $L_1$, \dots, $L_k$ be elements of~$\K[\x]\langle \Dx\rangle$,  with respective total degrees $\Delta_1$, \dots, $\Delta_k$.
For~$N\geq\max\{\Delta_1,\dots,\Delta_k\}$, let $M'_N(L_1,\dots,L_k)$ be the matrix
\vspace{-0.3cm}
\begin{equation*}
M'_N \; = \; M'_N(L_1,\dots,L_k) \; = \; 
\begin{pmatrix}
C_N(L_1) & & \\
 & \ddots & \\
 & & C_N(L_k) \\
-I_{\binom{N+2}2} & \dots & -I_{\binom{N+2}2}
\end{pmatrix}.
\end{equation*}
This matrix has $m(N)$~rows and~$n(N)$~columns, where
\[
m(N) \!=\! \binom{N+2}2 + \sum_{j=1}^k \binom{N-\Delta_j+2}2, \,
n(N) \!=\! k\binom{N+2}2.
\]

Assuming all~$\Delta_i$ equal to a same value~$\Delta$, the matrix~$M'_N$ certainly has a nontrivial left kernel when~$m(N)>n(N)$, that is when $ k\binom{N-\Delta+2}2 > (k-1)\binom{N+2}2,$
which happens when
\begin{equation*}
N \geq B
\;\text{for}\;
B = \left\lceil
  k\Delta+\frac{\sqrt{4k(k-1)\Delta^2+1}-3}2
\right\rceil \leq 2k\Delta,
\end{equation*}
where the approximation holds for large values of $k$ or~$\Delta$.

\smallskip Using  $\Delta \leq d+r$ yields the main result of this section.

\vspace{-0.3cm}
\begin{theorem} \label{th:CLM}
Let~$L_1, \ldots, L_k$ be elements in~$\K[\x]\langle \Dx\rangle \setminus \{
0\}$ of orders at most~$r$, and with coefficients of degrees at most~$d$.
There exist nonzero common left multiples of total degree~$\leq 2k(d+r)$ in
$(x,\partial)$, and total arithmetic size $O\bigl(k^2 (d+r)^2\bigr)$.
\end{theorem}

\vspace{-0.3cm}
\medskip \noindent {\bf Algorithms for CLMs.} A simple algorithm for computing  common left multiples of total degree~$B\approx 2k(d+r)$ in $(x,\partial)$ is  based on the left kernel
computation of the scalar matrix~$M'_{B}(L_1,\dots,L_k)$. This matrix has sizes of order $k
B^2/2 \approx 2k^3(d+r)^2$. The cost of the procedure is $O(k^{3\theta} (d+r)^{2\theta})$; it is dominated by the kernel computation.

\smallskip To simplify the discussion, we assume in the remaining of the
section that $L_1,\ldots,L_k$ have bidegrees at most $(d,r)=(n,n)$. 
Then, Theorem~\ref{th:CLM} implies that, while the LCLM has
order at most $kn$ and degrees at most $k^2n^2$, there exist common left
multiples of order and degree at most $4kn$.
However, computing such a small multiple by the previous algorithm of
complexity $O(k^{3\theta} n^{2\theta})$ is more costly than computing the
LCLM by the last two algorithms in Figure~\ref{fig:complexity}. 

Here we briefly sketch a faster algorithm for computing a common left multiple
of order and degree at most $4kn$, based on Hermite-Pad\'e
approximation~\cite[Chapter~10]{Bostan03}. One determines series
solutions of the~$L_i$ at order~$O(k^2 n^2)$, takes a random linear
combination~$f$ of them, computes its first $4kn$ derivatives, and outputs a
Hermite-Padé approximant of $f, f', \ldots, f^{(4kn)}$ of type $(4kn, \ldots,
4kn)$. The dominant complexity is that of the Hermite-Padé step,
$\bigOsoft(k^{\theta+1} n^{\theta+1})$~\cite{Storjohann06}.

\medskip \noindent {\bf A Fast LCLM Heuristic.} 
As an interesting consequence of this fast CLM computation, we
deduce a very efficient heuristic for LCLMs, asymptotically
faster than all algorithms in Figure~\ref{fig:complexity}. It proceeds in 3 steps:
($i$) compute~$O(1)$ CLMs of order and degree at most $4kn$;
($ii$) take two random linear combinations with coefficients in $\K[x]$; ($iii$)
return their GCRD. The dominating steps are~($i$) and ($iii$). By using
Hermite-Padé approximants for step ($i$) and Grigoriev's
algorithm~\cite{Grigoriev90} combined with Theorem~\ref{theo:SV} for step~($iii$), the total complexity is
$\bigOsoft(k^{\theta+1} n^{\theta+1})$. This is nearly optimal, in view of
the LCLM size $k^3 n^3$. 
However, we are not yet able to turn this heuristic into a fully proved algorithm.

\end{section}


\vspace{-0.1cm}

\section{Experiments}

We implemented\footnote{All computer calculations were performed on a
Quad-Core Intel Xeon X5160 processor at 3GHz, with 8GB of RAM.} two variants
of our new algorithm in Magma V2.16-7~\cite{magma} and compared them with
Magma's built-in LCLM routine (command \verb+LeastCommonLeftMultiple+).

Some experimental results are summarised in Table~\ref{tab:Magma-lclms}. We
take as input $k=2$ \emph{random\/} operators in $\F_p[\x]\langle \Dx\rangle$,
each of bidegree~$(d,r) = (n,n)$ in $(x,\Dx)$, where $p$ is a medium-sized
prime and $n$ is of the form $\lceil 2^{j/2} \rceil$, for $2\leq j \leq 11$.
Column {\sf New} gives timings for the first variant of the new algorithm,
that uses Magma's built-in polynomial linear algebra solver (the \verb+Kernel+
routine), while column {\sf New+S} gives timings for the second variant, based
on our own high-level implementation of Storjohann's \emph{high-order lifting
algorithm}~\cite{Storjohann03}. Column $(N,D)$ displays the size~$N$ and the
degree~$D$ of the polynomial matrix dealt with by algorithms {\sf New} and
{\sf New+S}. The dominating part of these algorithms is the left kernel
computation for a polynomial matrix of size~$(N+1) \times N$ and degree~$D$.
The most time consuming part of~{\sf New+S} consists in $O(\log N)$ polynomial
matrix multiplications of size $N$ and degree $D$. To facilitate comparisons,
column $\MM(N,D)$ shows the total time taken by 10 products of random
polynomial matrices of size~$N$ and degree~$D$ over~$\F_p$. Finally, column
{\sf output size} displays the total arithmetic size of the computed LCLM,
that is, its number of coefficients in~$\F_p$.

Several conclusions can be drawn from
Table~\ref{tab:Magma-lclms}. First, Magma's LCLM tool exhibits an
\emph{exponential\/} arithmetic complexity behaviour (when passing from
bidegree $(n,n)$ to $(n+1,n+1)$, timings are multiplied by a factor close
to~$1.5$), but it is relatively efficient for small input sizes. Both variants
of the new algorithm are faster for $n\geq 10$, and {\sf New+S} gains a factor
65 for $n=23$, and almost 1300 for $n=46$.

Second, timings in column {\sf New} exhibit a practical complexity
proportional to~$n^5$, which is inherited from Magma's linear algebra solver
on polynomial matrices. In contrast, {\sf New+S} has a practical complexity
proportional to $n^{3.5}$ (but with a higher proportionality factor). This
good behaviour, closer to the theoretical complexity $\bigOsoft(n^{\theta+1})$
predicted by Theorem~\ref{th:main}, is inherited from Magma's very efficient
polynomial matrix multiplication, through Storjohann's algorithm.

Finally, timings in column {\sf New+S} grow nearly linearly in the
corresponding output sizes given in the last column, and these sizes match
exactly the sharp bounds in Theorem~\ref{th:main}. This experimentally
confirms that size bounds and worst-case complexity analyses predicted by our
theoretical results 
are reached in \emph{generic\/} cases.

\begin{table}[t]
\begin{scriptsize}
\tabcolsep2pt
\begin{center}
\begin{tabular}{r|ccccccc}
$n$ & \sf Magma's LCLM & \sf New &  \sf New+S & $(N,D)$& $\MM(N,D)$& \sf output size \\
\hline
 2 & 0.01 & 0.00 & 0.01 & (2,10) & 0.01 & 65 &  \\
 3 & 0.01 & 0.01 & 0.03 & (3,14) & 0.01 & 175 &  \\
 4 & 0.02 & 0.01 & 0.07 & (4,18) & 0.03 & 369 &  \\
 6 & 0.10 & 0.06 & 0.17 & (6,26) & 0.06 & 1105 &  \\
 8 & 0.49 & 0.19 & 0.54 & (8,34) & 0.15 & 2465 &  \\
12 & 6.84 & 0.91 & 1.37 & (12,50) & 0.41 & 7825 &  \\
16 & 49.24 & 3.48 & 4.93 & (16,66) & 0.91 & 17985 &  \\
23 & 718.02 & 20.51 & 11.09 & (23,94) & 2.60 & 51935 &  \\
32 & 9355.47 & 115.53 & 40.83 & (32,130) & 6.73 & 137345 &  \\ 
46 & 168434.66 & 791.01 & 130.40 & (46,186) &  21.51 & 402225 \\
\end{tabular}
\end{center}
\end{scriptsize}
\vskip-12pt
\begin{small}
\caption{Timings (in sec.) for LCLMs of $k=2$ random operators in $\F_p[\x]\langle
\Dx\rangle$ of bidegrees $(n,n)$ in $(x,\Dx)$. }\label{tab:Magma-lclms} 
\end{small}
\end{table}

\scriptsize

\def\cprime{'} \def\cprime{$'$}

\end{document}